\documentclass[aps, prl, twocolumn, superscriptaddress, groupaddress, floatfix, longbibliography]{revtex4-1}
\usepackage{graphicx}
\usepackage{bm}
\usepackage{amssymb}
\usepackage{color}
\usepackage{amsmath, mathrsfs}
\usepackage{amstext}
\usepackage{float}
\usepackage{braket}
\usepackage{cancel}
\usepackage{oubraces}
\usepackage{latexsym}
\usepackage[usenames,dvipsnames]{xcolor}
\usepackage[colorlinks=true,citecolor=Blue,linkcolor=RubineRed,urlcolor=Blue]{hyperref}
\usepackage{bbold}
\usepackage{bbm}

\usepackage{stackengine}
\usepackage{graphicx}

\DeclareFontFamily{OT1}{pzc}{}
\DeclareFontShape{OT1}{pzc}{m}{it}{<-> s * [0.900] pzcmi7t}{}
\DeclareMathAlphabet{\mathpzc}{OT1}{pzc}{m}{it}

\def\H{\hat{H}}

\def\s{\hat{\sigma}}
\def\sd{\hat{\sigma}^{\dagger}}

\usepackage[colorinlistoftodos]{todonotes}

\begin{document}
\title{Time-Varying Medium from an Ensemble of Laser-Dressed Quantum Emitters}

\author{Adri\'an Juan-Delgado}
\email[]{adrianjuand1996@gmail.com}
\affiliation{Departamento de Física Teórica de la Materia Condensada, Universidad Autónoma de Madrid, E-28049 Madrid, Spain}
\affiliation{Condensed Matter Physics Center (IFIMAC), Universidad Autónoma de Madrid, E-28049 Madrid, Spain}

\author{Diego Martín-Cano}
\email[]{diego.martin.cano@uam.es}
\affiliation{Departamento de Física Teórica de la Materia Condensada, Universidad Autónoma de Madrid, E-28049 Madrid, Spain}
\affiliation{Condensed Matter Physics Center (IFIMAC), Universidad Autónoma de Madrid, E-28049 Madrid, Spain}
\affiliation{Instituto Nicolás Cabrera (INC), Universidad Autónoma de Madrid, E-28049 Madrid, Spain}

\author{Paloma A. Huidobro}
\email[]{p.arroyo-huidobro@uam.es}
\affiliation{Departamento de Física Teórica de la Materia Condensada, Universidad Autónoma de Madrid, E-28049 Madrid, Spain}
\affiliation{Condensed Matter Physics Center (IFIMAC), Universidad Autónoma de Madrid, E-28049 Madrid, Spain}
\affiliation{Instituto Nicolás Cabrera (INC), Universidad Autónoma de Madrid, E-28049 Madrid, Spain}

\date{\today}

\begin{abstract}
We propose a minimal quantum-mechanical model of a frequency dispersive, time-modulated medium that explicitly incorporates the pumping mechanism. It consists of a homogeneous ensemble of quantum emitters dressed by a pump laser. We analyze their dispersive, time-varying response to a perturbative probe field by deriving the macroscopic polarizability of the ensemble, and then solving Maxwell’s equations via the Floquet formalism. This yields a rich band structure that departs from conventional analyses in which the modulation source is neglected, arising from the interplay between the dispersive polaritonic and time-varying responses. These results shed light on the role of the modulation source in time-varying media and pave the way toward realizing dispersive time-varying media from driven quantum-mechanical systems.

\end{abstract} 

\maketitle

Photonic time-varying media (TVM), materials whose optical properties are modulated in time \cite{Zurita_PRA_2009, Galiffi_Review_2022, Carsten_tutorial_2024, Boltasseva_OME_2024}, have emerged as a promising platform for controlling light, enabling linear frequency conversion, amplification, and non-reciprocity. A hallmark of TVM is the emergence of momentum gaps in the photonic dispersion relation under fast periodic modulation \cite{Reyes_APL_2015,Galiffi_PRL_2020,Park_SciAd_2022, Gaxiola_APL_2023, Wang_SA_2023, Horsley_PRL_2023,Galiffi_NAtPhot_2026,Huang_arxiv_2026,Zhu_arxiv_2025}. Such momentum gaps host complex-frequency modes and can temporally amplify light when the modulation is sufficiently strong. However, theoretical analyses typically rely on classical electromagnetism, leaving the quantum-mechanical description of TVM largely unexplored \cite{Lyubarov_Science_2022, Sloan_ACSPhotonics_2024, Bae_NatComms_2025, Park_PRL_2025, Sustaeta_PRL_2026, Kim_COP_2026}, particularly from a miscrocopic point of view of the medium. Moreover, these analyses often assume a non-dispersive, phenomenological time-dependent permittivity, without explicitly accounting for the pumping mechanism responsible for the modulation. This raises an important question: how are conventional TVM effects modified when the quantum-mechanical nature of the medium, its dispersive response, and the underlying pump are all explicitly accounted for?

\begin{figure}[!t] 
	\begin{center}
		\includegraphics[width=0.483\textwidth]{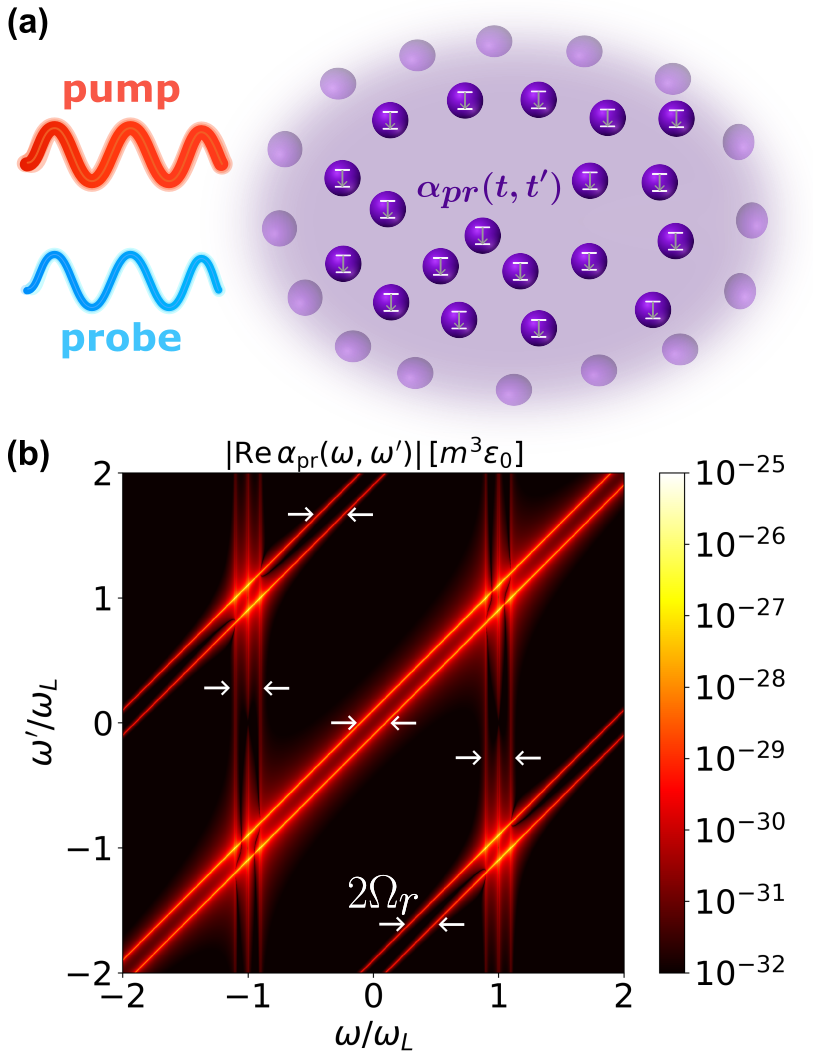}
		\caption{Polarizability for the minimal quantum-mechanical model of TVM proposed in this Letter. (a) Schematic representation of the homogeneous ensemble of QEs driven by a strong pump laser. The dispersive, time-varying response to a weak probe laser is characterized by the two-time polarizability $\alpha_{pr}(t,t')$. (b) Behaviour of polarizability in the frequency domain under resonant pumping. $|\text{Re}\:\alpha_{pr}(\omega, \omega')|$ is represented in units of $\varepsilon_0 m^{3}$. We fix $\Omega=\Omega_r=\omega_0 /10$. White arrows denote frequency splittings of value $2\Omega_r$.}   \label{Fig:1} 
	\end{center}
\end{figure}

In this Letter, we address this question by proposing a minimal quantum-mechanical model of a dispersive time-varying medium that explicitly includes the pumping mechanism. The model consists of a homogeneous ensemble of quantum emitters (QEs) driven into the Mollow regime [Fig.~\ref{Fig:1}(a)], where the pump laser dresses the states of the system \cite{Mollow_PR_1969,Schuda_JPB_1974,Walther_Proceedings_1975,Wu_PRL_1975}. We study the linear response of this driven system to a weak probe field by analytically deriving its dispersive polarizability, following the approach of Ref.~\cite{Sloan_ACSPhotonics_2024} but explicitly accounting for the pumping mechanism. This polarizability is inserted into Maxwell’s equations, which are then solved using the Floquet theorem to obtain the electromagnetic modes supported by the driven medium. The resulting photonic band structure, centered at the pump laser frequency, displays flat momentum gaps at the emitter dressed frequencies, with exceptional points at the band edges that we characterize via the phase rigidity \cite{Wiersig_PRR_2023,Allard_PRL_2026}, as well as dispersive momentum gaps and avoided crossings. These features are a direct consequence of including dispersion and the modulation source in the analysis.

We start by analyzing the dynamics of a single QE pumped by a laser, which will serve to build the macroscopic response from the ensemble of driven QEs. The ground and excited states of the QE are $\ket{g}$ and $\ket{e}$, and the transition frequency is $\omega_{0}$. 
Within the Rotating-Wave Approximation, the Hamiltonian of the pumped QE is \cite{Steck_book_2007, JuanDelgado_Thesis,Gerry_book_2005,Loudon_book_2000}
\begin{equation} \label{Eq:H_lab_frame}
    \H (t) = \hbar \omega_{0} \s_+ \s_- + \hbar \frac{\Omega}{2}(\s_- e^{i\omega_L t} + \s_+ e^{-i\omega_L t} ),
\end{equation}
where the first term on the right-hand side corresponds to the bare QE Hamiltonian and the second term accounts for the interaction of the QE with the pump field. Here, $\s_-=\ket{g}\bra{e}$ represents the lowering operator of the QE, $\s_+=\ket{e}\bra{g}$ is the raising operator, and $\Omega$ is the Rabi frequency. 
We neglect any losses because we aim to analyze the time-periodic response of the system, and thus focus on timescales short compared to the decay rates of the system. The unitary time evolution of the system is thus governed by the propagator $\hat{U}(t)=\hat{\mathcal{T}}\exp[-i\int_0^t dt' \H(t')]$ \cite{JuanDelgado_PRA_2021}, with $\hat{\mathcal{T}}$ the temporal ordering operator. 

The evaluation of the time evolution of the pumped system is facilitated by moving to the rotating frame at the laser frequency $\omega_L$ through the unitary transformation $\hat{P}(t)=\exp(-i\omega_L \sd\s)$, which yields the time-independent Hamiltonian $\H_R =\hbar \Delta \sd \s +\hbar \frac{\Omega}{2}(\s + \sd)$ in the rotating frame \cite{Steck_book_2007, JuanDelgado_Thesis}, where $\Delta=\omega_0 - \omega_L$ is the laser detuning. As a result, the propagator that governs the time evolution can be decomposed as $\hat{U}(t)=\hat{P}(t)\hat{R}(t)$, with $\hat{R}(t)=\exp(-i \H_R t)$, which does not involve any time-integral. This decomposition can be interpreted as a Floquet decomposition, since the Hamiltonian $\H (t)$ is time-periodic \cite{JuanDelgado_PRA_2021,Shirley_PR_1965,Zeldovich_JETP_1967,Szczygielski_LAA_2021}. Diagonalizing $\H_R$ yields the dressed states in the rotating frame,
\begin{subequations}
\begin{align}
    \ket{\Omega_+}&=\cos\theta \ket{e} + \sin\theta \ket{g}, \\
    \ket{\Omega_-}&=-\sin\theta \ket{e} + \cos\theta \ket{g},
\end{align}
\end{subequations}
where $\theta$ is the St\"uckelberg (mixing) angle \cite{Steck_book_2007,JuanDelgado_Thesis}, defined by $\cos(2\theta)=\Delta/\Omega_{r}$ and $\sin(2\theta)=\Omega/\Omega_{r}$. Here, $\hbar\Omega_{r} = \hbar\sqrt{\Omega^2 + \Delta^2}$ is the generalized Rabi frequency, which corresponds to the energy splitting between $\ket{\Omega_+}$ and $\ket{\Omega_-}$. In the Mollow regime, where the Rabi frequency dominates over the losses, the populations of the ground and excited bare states oscillate at frequency $\Omega_{r}$ \cite{Cohen_JPB_1977,Dalibard_JOSAB_1985,DelValle_PRL_2010,JuanDelgado_OPA_2026} -- known as Rabi oscillations -- and the emission spectrum displays the characteristic Mollow triplet: a central peak at the pump frequency $\omega_L$ and two sidebands at $\omega_L\pm\Omega_{r}$ \cite{Mollow_PR_1969,Schuda_JPB_1974,Walther_Proceedings_1975,Wu_PRL_1975}.

Our objective is to understand the electromagnetic response of such driven quantum system under a weak probe field. For that purpose, we derive the Heisenberg evolution of the dipole operator, which will be used to obtain the polarizability via the Kubo formula \cite{Kubo_JPSJ_I_1957,Kubo_JPSJ_II_1957}. For simplicity, we assume that the pump field and the transition dipole moment are linearly polarized along the same direction, which we fix along the $x$-axis. This allows us to work with the scalar dipole operator $\hat{d}=d(\hat\sigma_-+\hat\sigma_+)$, where $d$ is the magnitude of the transition dipole moment. Computing $\hat{d}(t)=\hat{U}^\dagger (t) \hat{d} \hat{U}(t)$, with $\hat{U}(t)=\hat{P}(t)\hat{R}(t)$, one obtains \cite{JuanDelgado_PRA_2021}
\begin{equation} \label{Eq:dipole_Heisenberg}
    \hat{d}(t)=\sum_{m,l\in\Lambda_1} e^{i(m\Omega_r + l\omega_L)t}d_{m,l}\hat{s}_m ,
\end{equation}
where we have introduced $\Lambda_1=\{-1,0,+1\}$ and the operators $\hat{s}_{-1}=\ket{\Omega_-}\bra{\Omega_+}$, $\hat{s}_{+1}=\ket{\Omega_+}\bra{\Omega_-}$, and $\hat{s}_{0}=\ket{\Omega_+}\bra{\Omega_+}-\ket{\Omega_-}\bra{\Omega_-}$, which are the lowering, raising and z-Pauli matrices in the basis $\{\ket{\Omega_+},\ket{\Omega_-}\}$. Additionally, $d_{\pm1,\pm1}=d(\cos(2\theta)+1)/2$, $d_{\pm1,\mp1}=d(\cos(2\theta)-1)/2$, $d_{0,\pm1}=d\sin(2\theta)/2$, and $d_{m,0}=0$.

We are now in a position to derive the two-time polarizability of a single driven emitter in response to a perturbative probe field. For simplicity, we consider that the probe electric field is also polarized in the $x$-direction. Within linear-response theory, the probe-induced polarizability is given by the Kubo formula \cite{Sloan_ACSPhotonics_2024}, $\alpha_{pr}(t,t')=\frac{i}{\hbar}\Theta(t-t')\braket{[\hat{d}(t),\hat{d}(t')]}$, where $\Theta(t-t')$ is the Heaviside step function and the expectation value is evaluated with respect to the initial state, which hereafter we fix to be the ground state $\ket{g}$. Using Eq. (\ref{Eq:dipole_Heisenberg}), we obtain
\begin{equation} \label{Eq:alpha_time}
    \begin{split}
        \alpha_{pr} &(t,t')=\frac{i\Theta(t-t')}{\hbar}\sum_{m,m',l,l'\in\Lambda_1 }\alpha_{m,l,m',l'} \\
        &\times \text{exp}[{i(m\Omega_{r} + l\omega_L)t}]\text{exp}[{i(m'\Omega_r + l'\omega_L)t'}],
    \end{split}
\end{equation}
where 
\begin{equation} \label{Eq:alpha_mlmplp_t}
    \begin{split}
        \alpha_{m,l,m',l'} &=d_{m,l}d_{m',l'} (\bra{g}\hat{s}_m\hat{s}_{m'}\ket{g}-\bra{g}\hat{s}_{m'}\hat{s}_{m}\ket{g}).
    \end{split}
\end{equation}
Equation (\ref{Eq:alpha_time}) reveals that $\alpha_{pr}(t,t')$ exhibits oscillatory dependence on both $t$ and $t'$, with contributions at the pump-laser and generalized Rabi frequencies. 

To further characterize $\alpha_{pr}$, we investigate its behaviour in the frequency domain via Fourier transformation of Eq. (\ref{Eq:alpha_time}). Figure \ref{Fig:1}(b) shows the absolute value of $\text{Re}\,\alpha_{pr}(\omega,\omega')$, for a resonantly driven QE ($\Delta=0$), with $\hbar\omega_L = 10 \Omega = 1$ eV. We choose this value of the Rabi frequency for the sake of visibility, as the effect is scalable to any value of $\Omega$ provided it remains much larger than the losses.
From Fig. \ref{Fig:1}(b), we identify two 
distinct spectral features. First, two manifolds of vertical lines, each split into three lines, which arise due to the dispersive response. Second, three manifolds of diagonal lines, 
separated by $2\omega_L$, each split into two lines 
separated by $2\Omega_r$ (white arrows). These diagonal lines correspond to fixed values of $\omega - \omega'$, revealing that the temporal modulation of the medium couples frequency components separated by some combination of $\omega_L$ and $\Omega$, a signature of a time-varying response \cite{Mirmoosa_NJP_2022,Verde_PRR_2026}. Analytical expressions for $\alpha_{pr}(\omega,\omega')$ are derived in the Supplemental Material (SM) \footnote{See Supplemental Material, which includes Refs. \cite{JuanDelgado_PRA_2021,Sloan_ACSPhotonics_2024,Galiffi_Review_2022,Carsten_tutorial_2024}, for the complete derivation of the analytical expression of the polarizability; the complete derivation of the Floquet eigenvalue problem with $\omega_F$ as eigenvalue, for the case in which $\omega_L$ is an integer multiple of $\Omega_r$; the extension of this eigenvalue problem to the general case in which $\omega_L$ is not an integer multiple of $\Omega_r$; an alternative derivation of the Floquet eigenvalue problem with $k^2$ as eigenvalue; and the analysis of the band structure for different values of $N_d$.}.

The electromagnetic response of a laser-dressed QE can be next extended to the macroscopic response of an homogeneous ensemble by assuming that all QEs are identical and non-interacting. The macroscopic polarization induced by the probe field is then obtained by multiplying the single-emitter polarizability $\alpha_{pr}$ by the emitter number density per unit volume $N_d$. 

In the following, we use Maxwell's equations and the Floquet theorem to derive the dispersion relation of the electromagnetic modes supported by the pumped medium that can be excited by a perturbative probe field. To this end, we consider that these modes propagate in the $z$-direction, with electric field, electric displacement field, magnetic field and polarization given by $\mathbf{E}(z,t)=E(t)e^{ikz}\mathbf{e}_x$, $\mathbf{D}(z,t)=D(t)e^{ikz}\mathbf{e}_x$, 
$\mathbf{B}(z,t)=B(t)e^{ikz}\mathbf{e}_y$, and 
$\mathbf{P}(z,t)=P(t)e^{ikz}\mathbf{e}_x$. The time evolution of these fields is governed by the Maxwell equations, together with the constitutive relation. Specifically, the Faraday and Ampère equations, respectively, yield
\begin{subequations} \label{Eq:Maxwell_Eqs}
\begin{align}
    \frac{\partial B (t)}{\partial t} = -ikE(t) , \label{Eq:Faraday} \\
    \frac{\partial D (t)}{\partial t} = -\frac{ik}{\mu_0}B(t), \label{Eq:Ampere}
\end{align}
\end{subequations}
with $\mu_0$ the magnetic vacuum permeability. Considering a vacuum background for the QEs, the constitutive relation is given by $D(t)=\varepsilon_0 E(t) + P(t)$, where the polarization can be obtained from the polarizability as
\begin{equation} \label{Eq:polarization_general}
\begin{split}
    P(t)&=N_d \int_{-\infty}^\infty dt'\: \alpha_{pr}(t,t')E(t').
\end{split}
\end{equation}
From this polarizability and the Maxwell's equations in Eq. (\ref{Eq:Maxwell_Eqs}), the dispersion relation of these modes can be obtained.

To simplify the analytical treatment and the numerical calculation, we henceforth assume that the laser frequency is an integer multiple of the generalized Rabi frequency, $\omega_L = n_0 \Omega_r$. The exponential $\text{exp}[i(m\Omega_r + l\omega_L)t]$ appearing in $\alpha_{pr}(t,t')$ [Eq. (\ref{Eq:alpha_time})] then reduces to $\text{exp}[in\Omega_r t]$, where $n=m+n_0 l$, reducing the number of indices from two to one. The polarization can thus be written as 
\begin{equation}
    P(t) = \frac{i }{\hbar}\sum_{n,n'} \varepsilon_{n,n'}e^{in\Omega_r t} Q^{(n')} (t) ,
\end{equation}
where we have defined 
\begin{equation}
    \varepsilon_{n,n'}=N_d \sum_{l,l'} (\alpha_{n-ln_0 ,l,n'-l'n_0, l'}-\alpha_{n'-l'n_0, l',n-ln_0 ,l}),
\end{equation}
as well as the auxiliary field variables
\begin{equation}
    Q^{(n)} (t) = \int_{-\infty}^\infty dt' \: \Theta(t-t')e^{in\Omega_r t'}E(t').
\end{equation}
This auxiliary variable satisfies the differential equation
\begin{equation} \label{Eq:differential_eq_Q}
    \frac{\partial Q^{(n)} (t)}{\partial t}=e^{in\Omega_r t}E(t),
\end{equation}
and plays a key role in constructing the Floquet eigenvalue problem, as they will serve to obtain a closed system of equations for the Floquet harmonics of the $D(t)$ and $B(t)$ fields in this time-varying, dispersive setup.

According to Floquet theory \cite{Zurita_PRA_2009, Carsten_tutorial_2024}, since the response function is periodic in time, we adopt the following harmonic expansions for the displacement field, the magnetic field, and the auxiliary variables:
\begin{subequations}
    \begin{align}
         B (t) = e^{-i\omega_F t} \sum_p B_p e^{ip\Omega_r t} , \\
         D (t) = e^{-i\omega_F t} \sum_p D_p e^{ip\Omega_r t} , \\
         Q^{(n)} (t) = e^{-i\omega_F t} \sum_p Q_p^{(n)} e^{ip\Omega_r t} ,
    \end{align}
\end{subequations}
where $\omega_{F}$ is the Floquet quasifrequency. Substituting these expansions into Eq. (\ref{Eq:Faraday}), Eq. (\ref{Eq:Ampere}), and Eq.~(\ref{Eq:differential_eq_Q}), we obtain 
\begin{subequations} \label{Eq:eigenvalue_problem}
    \begin{align}
         &\omega_F D_n =n\Omega_r D_n + \frac{k}{\mu_0}B_n ,\\
         &\omega_F B_n =n\Omega_r B_n + \frac{k}{\varepsilon_0}D_n - \frac{ik}{\varepsilon_0 \hbar} \sum_{n',p}\varepsilon_{n-p,n'}Q_{p}^{(n')} , \\
        & \omega_F  Q_{n'}^{(n)}= n'\Omega_r Q_{n'}^{(n)}+\frac{i}{\varepsilon_0}D_{n'-n}+\sum_{p,p'}\frac{\varepsilon_{n'-p',p}}{\hbar\varepsilon_0}Q_{p'-n}^{(p)}.
    \end{align}
\end{subequations}
These coupled equations constitute an eigenvalue problem in the Floquet quasifrequency, whose solution yields the allowed $(k,\omega_F)$ modes. In the SM \cite{Note1}, we describe the extension of this formulation to the general case in which $\omega_L$ is not necessarily an integer multiple of $\Omega_r$.

\begin{figure}[!t]  
	\begin{center}
		\includegraphics[width=0.48\textwidth]{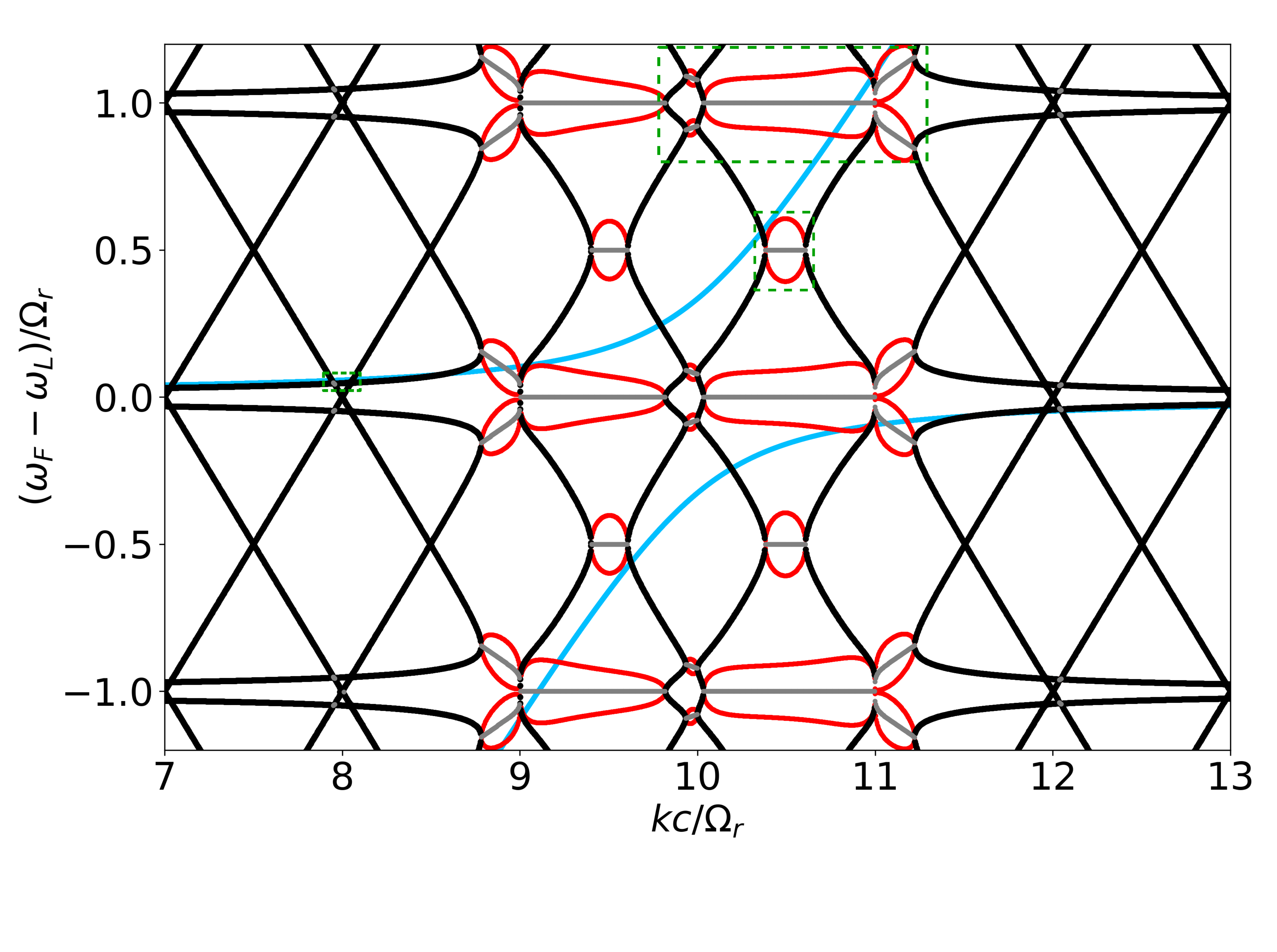}%
		\caption{Dispersion relation for a homogeneous ensemble of laser-dressed QEs. Modes with purely real Floquet frequency are represented by black dots, with the $y$-axis value corresponding to $(\text{Re}\{\omega_F\}-\omega_L)/\Omega_r$. Modes with a complex Floquet frequency are represented by grey and red dots: the grey dots show $(\text{Re}\{\omega_F\}-\omega_L)/\Omega_r$, while the red dots show the value of $\text{Im}\{\omega_F\}/\Omega_r$ shifted by $(\text{Re}\{\omega_F\}-\omega_L)/\Omega_r$. Blue dots correspond to the dispersion relation in the absence of a pump. The emitter and laser parameters are the same as in Fig.~\ref{Fig:1}(b), and $N_d=3\times10^{23}$ QEs per $\text{m}^{3}$. Green dashed boxes indicate the regions analyzed in Fig. \ref{Fig:rigidity}.} \label{Fig:Bands}
	\end{center}
\end{figure}

In the remainder of this Letter, we numerically solve the eigenvalue problem in Eq.~(\ref{Eq:eigenvalue_problem}) and analyze the resulting band structure. Specifically, we plot in Fig.~\ref{Fig:Bands} the band structure obtained for $N_d = 3\times 10^{23}$ QEs per $\text{m}^3$, which is sufficient to observe the emergence of momentum gaps and consistent with experiments \cite{Acosta_PRB_2009,Manson_NJP_2018,Auzinsh_PRB_2019,Skoff_PRA_2011,Asano_JAP_2010}. The remaining emitter and laser parameters are as in Fig.~\ref{Fig:1}(b) \footnote{To determine the value of the dipole moment, we use the spontaneous emission formula $d=\sqrt{3\pi\gamma\varepsilon_0\hbar c^3/\omega_0^3}$, which connects the dipole moment $d$ of a two-level QE to its transition frequency $\omega_0$ and decay rate $\gamma$. We set $\hbar\gamma=10^{-7}$ eV, a value representative of the losses in organic molecules at cryogenic temperatures, which also justifies neglecting dissipation in the derivation of the Floquet eigenvalue problem.}. Black dots denote modes with purely real Floquet frequency, grey dots denote the real frequency part of modes with complex Floquet frequency, and red dots denote the corresponding imaginary part, shifted vertically by its real part for clarity. We focus on the range of Floquet frequencies around the pump frequency $\omega_L=\omega_0$, in contrast with conventional analyses of periodically time-varying media, where the first Brillouin zone is centered at zero frequency. The band structure is thus centered precisely within the range where the Rabi splitting of the undriven polaritonic dispersion falls \cite{Kramers_1925, Loudon_JOPB_2006, Berman_PRA_2006}. To aid the interpretation of the driven band structure, we overlay in Fig.~\ref{Fig:Bands} the dispersion relation in the undriven case (blue dots) \footnote{This dispersion relation is given by $k = \frac{\omega_F}{c}\sqrt{1+\frac{N_d}{\varepsilon_0}\alpha_0(\omega)}$, with $\alpha_0(\omega)=\frac{d^2}{\hbar}\!\left[\frac{1}{\omega+\omega_0}-\frac{1}{\omega-\omega_0}\right]$}, which exhibits the characteristic upper and lower polariton branches separated by the Rabi splitting at $k = \omega_0/c = 10\Omega_r/c$. This polaritonic structure underlies the dispersive band structure of the driven medium.

\begin{figure}[!t]  
	\begin{center}
		\includegraphics[width=0.45\textwidth]{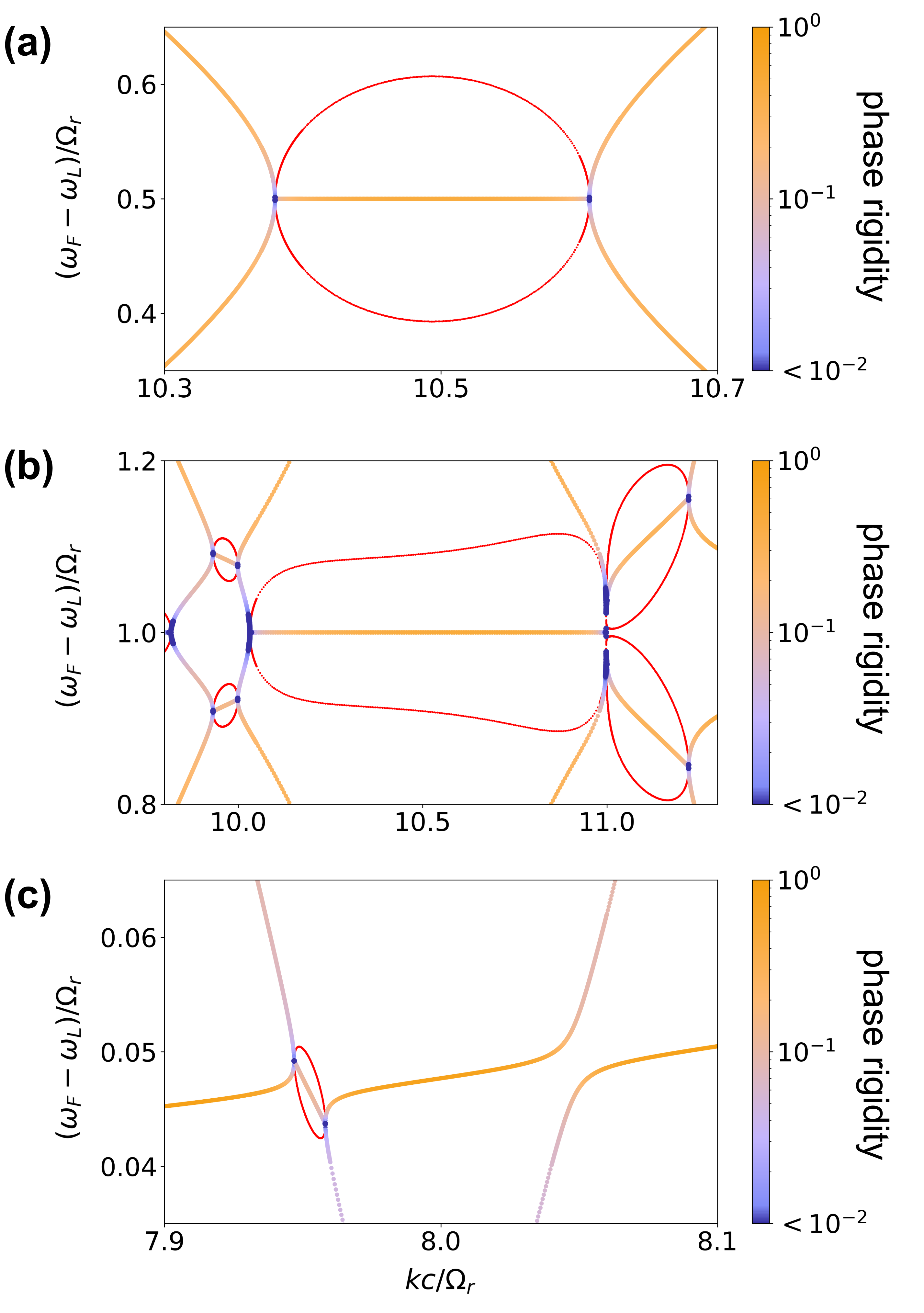}%
		\caption{Phase rigidity around the momentum gaps. The color bar indicates the value of the phase rigidity, with purple indicating values close to $0$. Red dots show the value of $\text{Im}\{\omega_F\}/\Omega_r$ shifted by $(\text{Re}\{\omega_F\}-\omega_L)/\Omega_r$. The emitter and laser parameters are the same as in Fig.~\ref{Fig:Bands}.} \label{Fig:rigidity}
	\end{center}
\end{figure}
The driven band structure in Fig.~\ref{Fig:Bands} organizes into two physically distinct regions in momentum space. In the $k$-region centered around the pump momentum ($9\Omega_r/c \lesssim k \lesssim 11\Omega_r/c$), the coupling between positive- and negative-frequency Floquet replicas systematically gives rise to momentum gaps. In the outer $k$-region ($kc/\Omega_r \lesssim 9$ and $kc/\Omega_r \gtrsim 11$), the coupling is weaker and momentum gaps do not generally open. We discuss these two regions in the following.

In the $k$-region around the pump momentum, positive- and negative-frequency Floquet replicas interact giving rise to several momentum gaps of different origin. First, flat momentum gaps open at $\mathrm{Re}(\omega_F) = \omega_L \pm \Omega_r /2$, arising from the parametric coupling of two photonic modes whose frequencies differ by $2\omega_L\pm\Omega_r$ \cite{Holberg_IEEE_1966} -- for example, modes $\exp[ikz - i(\omega_F + \omega_L)]$ and $\exp[ikz + i(\omega_F + \omega_L - \Omega_r)]$. These gaps are the hallmark of parametric amplification of the probe assisted by the pump, resembling the conventional momentum gaps \cite{Zurita_PRA_2009, Galiffi_Review_2022, Carsten_tutorial_2024, Boltasseva_OME_2024}. To further characterize these momentum gaps, we compute the phase rigidity for each mode $r_{\perp}=|\braket{L|R}|/\sqrt{\braket{L|L}\braket{R|R}}\in[0,1]$ \cite{Wiersig_PRR_2023,Allard_PRL_2026}, where $\bra{L}$ and $\ket{R}$ are the left and right eigenvectors of the Floquet matrix [Eq.~(\ref{Eq:eigenvalue_problem})] for this mode. A zoom of one of these flat gaps is shown in Fig.~\ref{Fig:rigidity}(a), where $r_\perp \to 0$ (purple in the color bar) marks exceptional points (EPs) at the band edges, consistent with the conventional case.

Second, dispersive momentum gaps appear at $\mathrm{Re}(\omega_F)\approx \omega_L$ and $\mathrm{Re}(\omega_F) \approx \omega_L \pm \Omega_r$, arising from the interaction between polaritonic and photonic modes of opposite frequency sign \cite{Feng_APL_2024, Ozlu_PRL_2025, Wang_APL_2025, Allard_PRL_2026}. A zoom of one of these regions is shown in Fig.~\ref{Fig:rigidity}(b). The smallest dispersive gaps connect the two sides of the central $k$-region around the pump momentum, and stem from the coupling of photonic modes to a next-order replica of the polaritonic mode. The largest dispersive gaps bridge each of these two sides of the central $k$-region with the outer $k$-regions. In addition, broad flat momentum gaps appear at $\mathrm{Re}(\omega_F) = \omega_L$ and $\mathrm{Re}(\omega_F) = \omega_L \pm \Omega_r$. We note that gain at Mollow sideband frequencies has also been predicted \cite{Mollow_PRA_1972} and observed experimentally \cite{Wu_Mollow_PRL_1977,Gruneisen_PRA_1989,Maser_NatPhot_2016} in laser-dressed quantum emitters, but in a steady-state regime where amplification manifests as spatial gain [$\mathrm{Im}(k) < 0$] rather than temporal gain [$\mathrm{Im}(\omega_F) > 0$]. EPs appear at the edges of all these gaps, as confirmed by the vanishing phase rigidity in Fig.~\ref{Fig:rigidity}(b).

In the outer $k$-region, momentum gaps are generally absent due to the weaker coupling between modes, as already mentioned. However, an exception arises when the polaritonic dispersion brings replicas of opposite frequency sign into resonance. This is illustrated in Fig.~\ref{Fig:rigidity}(c), which shows a zoom around $kc/\Omega_r \approx 8$: the intersection of the photonic and polaritonic bands gives rise either to a momentum gap with EPs at its edges (around $kc/\Omega_r = 7.95$, where the interacting replicas carry frequencies of opposite sign) or to an avoided crossing (around $kc/\Omega_r = 8.05$, where they carry frequencies of equal sign). Altogether, these results highlight the rich band structure that emerges within the minimal quantum-mechanical model of an ensemble of laser-dressed quantum emitters. 

In summary, we have derived the photonic band structure of a minimal quantum-mechanical model of a dispersive, time-varying medium that explicitly incorporates the source of the temporal modulation. The medium consists of a homogeneous ensemble of laser-dressed QEs, making our results applicable to a broad range of physical implementations, including organic molecules, atoms, quantum dots, and color centers. The resulting band
structure exhibits qualitative differences from that predicted by models assuming ad-hoc, non-dispersive modulations of the permittivity, due to the interplay between the dispersive polaritonic and time-periodic responses of the laser-dressed QEs, and could find applications in the engineering of momentum gaps \cite{Park_SciAd_2022,Wang_NatPhot_2025,Wang_APL_2025,Ozlu_PRL_2025,Caballero_ACS_2026,Allard_PRL_2026,Salehi_arxiv_2026,Huang_arxiv_2026}. This framework can serve to investigate light amplification and gain in driven systems beyond classical TVM treatments, shedding light on their potential as sources of quantum light. Next steps include a rigorous incorporation of losses, which may suppress the EPs of dispersive gaps without necessarily inhibiting amplification \cite{Allard_PRL_2026,LevelAttraction_PRA_2018}, and going beyond the homogeneous approximation.

\hfill

\begin{acknowledgments}
\textit{Acknowledgments}.--  We thank Thomas F. Allard, Miguel Verde, and Jaime E. Sustaeta-Osuna for helpful discussions. We have used AI tools (Claude) to optimize the numerical code used to solve the Floquet eigenvalue problem. A.J.D and P.A.H acknowledge funding from the European Union through the ERC Grant TIMELIGHT under GA101115792 and from the Spanish Ministry for Science, Innovation, and Universities - Agencia Estatal de Investigaci\'on (AEI) through Grant No. RYC2021-031568-I (Ram\'on y Cajal program). D.M.C. acknowledges support from MCINN projects PID2021-126964OB-I00 (QENIGMA), TED2021-130552B-C21 (ADIQUNANO), and PID2024-156077OB-I00 (DQUOTE), and the support by the project NPhoQuss funded by Programa Fundamentos 2024 from the BBVA Foundation through the grant EIC24-1-17304. We acknowledge funding from the Spanish Ministry for Science, Innovation, and Universities - Agencia Estatal de Investigaci\'on (AEI) through Grants No. CEX2023001316-M (through the Mar\'ia de Maeztu program for Units of Excellence in Research and Developments)  \\

\textit{Data availability}.-- The data that support the findings of this manuscript will be available online.
\end{acknowledgments}

\bibliographystyle{apsrev4-1}
\bibliography{bibliography} 
\end{document}


\title{Supplemental Material: Time-Varying Medium from an Ensemble of Laser-Dressed Quantum Emitters}

\author{Adri\'an Juan-Delgado}
\email[]{adrianjuand1996@gmail.com}
\affiliation{Departamento de Física Teórica de la Materia Condensada, Universidad Autónoma de Madrid, E-28049 Madrid, Spain}
\affiliation{Condensed Matter Physics Center (IFIMAC), Universidad Autónoma de Madrid, E-28049 Madrid, Spain}

\author{Diego Martín-Cano}
\email[]{diego.martin.cano@uam.es}
\affiliation{Departamento de Física Teórica de la Materia Condensada, Universidad Autónoma de Madrid, E-28049 Madrid, Spain}
\affiliation{Condensed Matter Physics Center (IFIMAC), Universidad Autónoma de Madrid, E-28049 Madrid, Spain}
\affiliation{Instituto Nicolás Cabrera (INC), Universidad Autónoma de Madrid, E-28049 Madrid, Spain}

\author{Paloma A. Huidobro}
\email[]{p.arroyo-huidobro@uam.es}
\affiliation{Departamento de Física Teórica de la Materia Condensada, Universidad Autónoma de Madrid, E-28049 Madrid, Spain}
\affiliation{Condensed Matter Physics Center (IFIMAC), Universidad Autónoma de Madrid, E-28049 Madrid, Spain}
\affiliation{Instituto Nicolás Cabrera (INC), Universidad Autónoma de Madrid, E-28049 Madrid, Spain}

\date{\today}

\maketitle

\onecolumngrid
\tableofcontents
\section{SI. Complete derivation of the polarizability} \label{Appendix:polarizability}
In this section, we provide further details on the analytical derivation of the expression presented in the main text for the time-domain polarizability $\alpha_{pr}(t,t')$ [Eq. (4) of the main text]. Additionally, we derive the analytical expression for the frequency-domain polarizability $\alpha_{pr}(\omega,\omega')$. 

Within linear-response theory, the time-domain polarizability is obtained via the Kubo formula \cite{Sloan_ACSPhotonics_2024}
\begin{equation} \label{Eq:App_Kuboformula}
    \alpha_{pr}(t,t')=\frac{i}{\hbar}\Theta(t-t')\braket{[\hat{d}(t),\hat{d}(t')]},
\end{equation}
where $\Theta(t-t')$ is the Heaviside step function, and $\hat{d}$ is the transition dipole moment operator. The expectation value is taken with respect to the initial state and the operators are evolved in the Heisenberg picture with respect to the bare Hamiltonian including the pump field (equivalently, in the interaction picture with respect to a probe field). To compute this expression, we begin from the Heisenberg equation of motion for the dipole moment operator $\hat{d}$ [Eq. (3) of the main text, which is derived in Ref. \cite{JuanDelgado_PRA_2021}],
\begin{equation} \label{Eq:App_dipole_Heisenberg}
    \hat{d}(t)=\sum_{m,l\in\Lambda_1} e^{i(m\Omega_r + l\omega_L)t}d_{m,l}\hat{s}_m ,
\end{equation}
where $\Lambda_1=\{-1,0,+1\}$, $\omega_L$ is the pump-laser frequency, and $\Omega_r = \sqrt{\Delta^2 + \Omega^2}$ is the generalized Rabi frequency, with $\Delta=\omega_0 - \omega_L$ the detuning between the emitter transition frequency and the pump-laser frequency. The operators $\hat{s}_{-1}=\ket{\Omega_-}\bra{\Omega_+}$.  $\hat{s}_{+1}=\ket{\Omega_+}\bra{\Omega_-}$, and $\hat{s}_{0}=\ket{\Omega_+}\bra{\Omega_+}-\ket{\Omega_-}\bra{\Omega_-}$, are, respectively, the lowering, raising and z-Pauli matrices in the basis $\{\ket{\Omega_+},\ket{\Omega_-}\}$ [see Eq. (2) of the main text]. The coefficients $d_{m,l}$ are given by $d_{\pm1,\pm1}=d(\cos(2\theta)+1)/2$, $d_{\pm1,\mp1}=d(\cos(2\theta)-1)/2$, $d_{0,\pm1}=d\sin(2\theta)/2$, and $d_{m,0}=0$, where $\cos(2\theta)=\Delta/\Omega_r$,  and $\sin(2\theta)=\Omega/\Omega_r$. We note that under resonant illumination ($\Delta=0$) and positive $\Omega$ (i.e., $\Omega_r=\Omega$), these coefficients reduce to $d_{\pm1,\pm1}=d/2$, $d_{\pm1,\mp1}=-d/2$, and $d_{0,\pm1}=d/2$, while the dressed states (in the rotating frame) are simply $\ket{\Omega_+}=(\ket{e}+\ket{g})/\sqrt{2}$ $\ket{\Omega_-}=(-\ket{e}+\ket{g})/\sqrt{2}$.

For an initial ground state $\ket{g}$, the two-time correlation function of the dipole moment operator becomes
\begin{equation}
\braket{d(t)d(t')}=\bra{g}d(t)d(t')\ket{g}=\sum_{m,m',l,l'\in\Lambda_1}e^{i(m\Omega_r + l\omega_L)t}e^{i(m'\Omega_r + l'\omega_L)t'}d_{m,l}d_{m',l'} \bra{g}\hat{s}_m\hat{s}_{m'}\ket{g} .
\end{equation}
The commutator in Eq. (\ref{Eq:App_Kuboformula}) is then given by
\begin{equation}
    \begin{split}
\braket{[d(t),d(t')]}&=\sum_{m,m',l,l' \in\Lambda_1}d_{m,l}d_{m',l'}\bra{g}\hat{s}_m\hat{s}_{m'}\ket{g} \biggr(e^{i(m\Omega_r + l\omega_L)t}e^{i(m'\Omega_r + l'\omega_L)t'}  - e^{i(m\Omega_r + l\omega_L)t'}e^{i(m'\Omega_r + l'\omega_L)t}\biggr).
    \end{split}
\end{equation}
Substituting into Eq. (\ref{Eq:App_Kuboformula}), the resulting expression of the time-domain polarizability is
\begin{equation}
    \begin{split}
        \alpha_{pr} (t,t')=\frac{i\Theta(t-t')}{\hbar}\sum_{m,m',l,l'\in\Lambda_1}d_{m,l}d_{m',l'} \bra{g}\hat{s}_m\hat{s}_{m'}\ket{g}\biggr(e^{i(m\Omega_r + l\omega_L)t}e^{i(m'\Omega_r + l'\omega_L)t'}  - e^{i(m\Omega_r + l\omega_L)t'}e^{i(m'\Omega_r + l'\omega_L)t}\biggr),
    \end{split}
\end{equation}
which can be written in the form of Eq. (4) of the main text by introducing
\begin{equation} \label{Eq:alpha_mlmplp_t}
    \begin{split}
        \alpha_{m,l,m',l'} &=d_{m,l}d_{m',l'} (\bra{g}\hat{s}_m\hat{s}_{m'}\ket{g}-\bra{g}\hat{s}_{m'}\hat{s}_{m}\ket{g}).
    \end{split}
\end{equation}

We next calculate the frequency-domain polarizability by computing the Fourier transform of $\alpha_{pr} (t,t')$, which is given by
\begin{equation} \label{Eq:App_Fourier_transform_polarizability}
    \alpha(\omega',\omega')=\int_{-\infty}^{\infty} dt\int_{-\infty}^{\infty} dt'e^{i(\omega t-\omega't')}\alpha(t,t').
\end{equation}
Using the integral representation of the Heaviside step function \cite{Sloan_ACSPhotonics_2024}, $\Theta(t-t')=\frac{i}{2\pi}\int_{\infty}^{\infty} d\omega'' \frac{e^{-i\omega'' (t-t')}}{\omega''+i\eta} $, where $\eta$ is a positive infinitesimal, we obtain that Eq. (\ref{Eq:App_Fourier_transform_polarizability}) becomes
\begin{equation}
    \begin{split}
        \alpha_{pr} (\omega,\omega')&=-\frac{1}{2\pi\hbar}\int_{\infty}^{\infty} d\omega'' \frac{e^{-i\omega'' (t-t')}}{\omega''+i\eta}\sum_{m,m',l,l'\in \Lambda_1}\bra{g}\hat{s}_m\hat{s}_{m'}\ket{g} d_{m,l}d_{m',l'} \\
        &\times \biggr[\int_{\infty}^{\infty} dt e^{i\omega t} e^{i(m\Omega_r + l\omega_L)t} \int_{\infty}^{\infty} dt'e^{-i\omega't'}e^{i(m'\Omega_r + l'\omega_L)t'} \\
        &- \int_{\infty}^{\infty} dt e^{i\omega t} e^{i(m'\Omega_r + l'\omega_L)t} \int_{\infty}^{\infty} dt'e^{-i\omega't'}e^{i(m\Omega_r + l\omega_L)t'}\biggr].
    \end{split}
\end{equation}
We next group the terms oscillating at $t$ and $t'$, which yields
\begin{equation}
    \begin{split}
        \alpha_{pr} (\omega,\omega')&=-\frac{1}{2\pi\hbar}\sum_{m,m',l,l' \in \Lambda_1}\bra{g}\hat{s}_m\hat{s}_{m'}\ket{g} d_{m,l}d_{m',l'} \int_{\infty}^{\infty} d\omega'' \frac{1}{\omega''+i\eta} \\
       &\times\biggr[\int_{\infty}^{\infty} dt e^{i(\omega - \omega'' + m\Omega_r + l\omega_L)t} \int_{\infty}^{\infty} dt'e^{i(-\omega' + \omega''+ m'\Omega_r + l'\omega_L)t'} \\
       &-\int_{\infty}^{\infty} dt e^{i(\omega - \omega'' + m'\Omega_r + l'\omega_L)t} \int_{\infty}^{\infty} dt'e^{i(-\omega' + \omega''+m\Omega_r + l\omega_L)t'} \biggr]. 
    \end{split}
\end{equation}   
Applying the identity $\delta(x)=\frac{1}{2\pi}\int_{-\infty}^\infty dt e^{itx}$, we obtain
\begin{equation}
    \begin{split}
        \alpha_{pr} (\omega,\omega')&=-\frac{1}{2\pi\hbar}\sum_{m,m',l,l'\in \Lambda_1}\bra{g}\hat{s}_m\hat{s}_{m'}\ket{g} d_{m,l}d_{m',l'} \int_{\infty}^{\infty} d\omega'' \frac{1}{\omega''+i\eta} \\
        &\times\biggr[2\pi\delta(\omega - \omega'' + m\Omega_r + l\omega_L)2\pi\delta(-\omega' + \omega''+ m'\Omega_r + l'\omega_L) \\
        &- 2\pi\delta(\omega - \omega'' + m'\Omega_r + l'\omega_L)2\pi\delta(-\omega' + \omega''+m\Omega_r + l\omega_L) \biggr].
     \end{split}
\end{equation}  
Evaluating next the integral over $\omega''$ via the Dirac delta functions leads to
\begin{equation}
    \begin{split}
        \alpha_{pr} (\omega,\omega')&=-\frac{2\pi}{\hbar}\sum_{m,m',l,l'\in \Lambda_1}\bra{g}\hat{s}_m\hat{s}_{m'}\ket{g} d_{m,l}d_{m',l'} \\
        &\times \biggr[ \frac{1}{\omega  + m\Omega_r + l\omega_L+i\eta} \delta(-\omega' + \omega  + m\Omega_r + l\omega_L+ m'\Omega_r + l'\omega_L) \\
        &-\frac{1}{\omega  + m'\Omega_r + l'\omega_L+i\eta}  \delta(-\omega' +\omega + m'\Omega_r + l'\omega_L+m\Omega_r + l\omega_L)\biggr] ,
    \end{split}
\end{equation}
which can be written in the compact form
\begin{equation} \label{Eq:App_alpha_freq_2_indices}
    \begin{split}
        \alpha_{pr} (\omega,\omega')=\sum_{m,m',l,l'\in \Lambda_1}\alpha_{m,m',l,l'}^{pr}(\omega)2\pi\delta(\omega -\omega' + (m+m')\Omega_r + (l+l')\omega_L) .
    \end{split}
\end{equation}
where we have defined 
\begin{equation}
    \alpha_{m,m',l,l'}^{pr}(\omega)=-\frac{d_{m,l}d_{m',l'}\bra{g}\hat{s}_m\hat{s}_{m'}\ket{g}}{\hbar}\biggr[ \frac{1}{\omega  + m\Omega_r + l\omega_L+i\eta} -\frac{1}{\omega  + m'\Omega_r + l'\omega_L+i\eta} \biggr].
\end{equation}
In the main text, Fig. 1(b) is obtained by using Eq. (\ref{Eq:App_alpha_freq_2_indices}), with $\hbar\eta=10^{-7}$ eV.

Interestingly, we can reduce the number of indices in Eq. (\ref{Eq:App_alpha_freq_2_indices}) from four to two, which facilitates the alternative derivation of the Floquet eigenvalue problem with $k^2$ as eigenvalue (Section SII of this Supplemental Material). To this end, we introduce $M=-m-m'$ and $L=-l-l'$, so that 
\begin{equation}
     \alpha_{pr} (\omega,\omega') = \sum_{M,L\in\Lambda_2} \alpha_{M,L}^{pr}(\omega)2\pi\delta(\omega -\omega' -M\Omega_r -L\omega_L) .
\end{equation}
We have introduced $\Lambda_2 = \{-2,-1,0,1,2\}$ and we have defined
\begin{equation} \label{Eq:App_alpha_ML}
    \alpha_{M,L}(\omega)=-\sum_{m,l}\frac{d_{m,l}d_{-M-m,-L-l}\bra{g}\hat{s}_m\hat{s}_{-M-m}\ket{g}}{\hbar}\biggr[ \frac{1}{\omega  + m\Omega_r + l\omega_L+i\eta} -\frac{1}{\omega  -(M+m)\Omega_r -(L+l)\omega_L+i\eta} \biggr].
\end{equation}
The summation over $m$ and $l$ in Eq. (\ref{Eq:App_alpha_ML}) is restricted to values for which $-M-m$ and $-L-l$ both belong to $\Lambda_1$. More specifically, if $M=2$ then $m$ runs only over $m=-1$; if $M=1$, over $m\in\{-1,0\}$; if $M=0$, over $m\in\Lambda_1$; if $M=-1$, over $m\in\{0,1\}$; and if $M=-2$, only over $m=1$. The same applies to $L$ and $l$.

\begin{figure}[!t] 
	\begin{center}
		\includegraphics[width=0.8\textwidth]{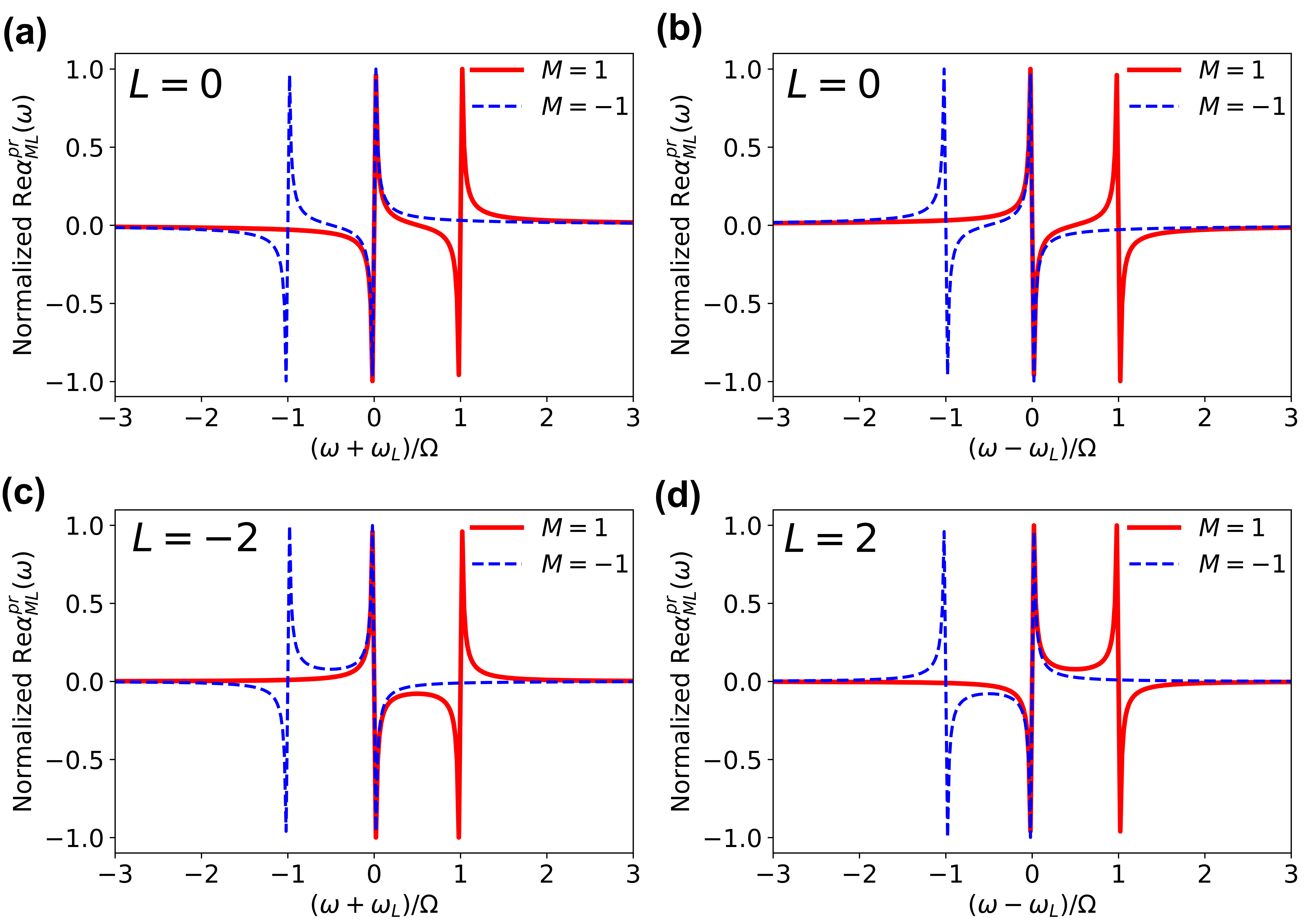}%
		\caption{Normalized Re$\alpha_{M,L}^{pr}(\omega)$ coefficients under resonant pumping. The $L$ index takes values (a,b) $L=0$, (c) $L=-2$, and (d) $L=2$. Red solid lines correspond to $M=1$ and dashed blue lines to $M=-1$. All other combinations of $(M,L)$ combinations give $\alpha_{M,L}^{pr}(\omega)=0$. We fix $\Omega=\Omega_r=\omega_0 /10$, $\hbar\omega_0 = 1$ eV, and $\hbar\eta=10^{-7}$ eV.}  
        \label{Fig:alpha_ML}  
	\end{center}
\end{figure}

To further characterize the behaviour of the polarizability, we plot in Fig. \ref{Fig:alpha_ML} the non-zero $\alpha_{M,L}(\omega)$ coefficients, normalized by $\max\{\text{Re}\,\alpha_{M,L}(\omega)\}$. Specifically, we find that $\alpha_{M,L}(\omega)$ is only non-zero for $M=1$ (solid red line) and $M=-1$ (dashed blue line), and for $L=-2,0,2$. Figures \ref{Fig:alpha_ML}(a) and \ref{Fig:alpha_ML}(b) show the real part of $\alpha_{M,L}(\omega)$ for $L=0$, associated with the antidiagonal central lines in Fig. 1(b) of the main text, plotted as a function of $(\omega+\omega_L)/\Omega$ and $(\omega-\omega_L)/\Omega$, respectively, so as to center the two relevant resonance pairs in each panel. In both cases, each curve displays two dispersive lineshapes with sharp sign transitions of width $\eta$. For $M=1$ (solid red line), $\alpha_{1,0}(\omega)$ exhibits these transitions at $\omega = -\omega_L$ and $\omega = -\omega_L + \Omega$ in panel (a), and at $\omega = \omega_L$ and $\omega = \omega_L + \Omega$ in panel (b). For $M=-1$ (dashed blue line), the transitions occur at $\omega = -\omega_L$ and $\omega = -\omega_L - \Omega$ in panel (a), and at $\omega = \omega_L$ and $\omega = \omega_L - \Omega$ in panel (b). Notably, the two curves for $M=\pm 1$ share one common resonance at $\omega = \pm\omega_L$, while the second resonance of each is displaced by $\pm\Omega$, reflecting the sideband structure induced by the drive.

Figure \ref{Fig:alpha_ML}(c) shows $\alpha_{M,-2}(\omega)$ for $L=-2$, plotted as a function of $(\omega+\omega_L)/\Omega$. For $M=1$ (solid red line), the sign transitions occur at $\omega = -\omega_L$ and $\omega = -\omega_L + \Omega$, and for $M=-1$ (dashed blue line) they appear at $\omega = -\omega_L$ and $\omega = -\omega_L - \Omega$. In contrast to the $L=0$ case, the two dispersive features within each curve now have equal amplitudes, so that the $M=1$ curve is symmetric about $\omega = -\omega_L + \Omega/2$ and the $M=-1$ curve is symmetric about $\omega = -\omega_L - \Omega/2$.

Figure \ref{Fig:alpha_ML}(d) shows $\alpha_{M,2}(\omega)$ for $L=2$, plotted as a function of $(\omega-\omega_L)/\Omega$. For $M=1$ (solid red line), the sign transitions occur at $\omega = \omega_L$ and $\omega = \omega_L + \Omega$, and for $M=-1$ (dashed blue line) they appear at $\omega = \omega_L$ and $\omega = \omega_L - \Omega$. As in panel (c), the two dispersive features within each curve have equal amplitudes, so that the $M=1$ curve is symmetric about $\omega = \omega_L + \Omega/2$ and the $M=-1$ curve is symmetric about $\omega = \omega_L - \Omega/2$.

\section{SII. Floquet eigenvalue problem with Floquet quasifrequency as eigenvalue} \label{Appendix:Floquet_frequency}
In this section, we derive in detail the Floquet eigenvalue problem with the Floquet quasifrequency as eigenvalue. We do so first by considering that the laser frequency is an integer multiple of the generalized Rabi frequency, as assumed in the main text, and then extend the formulation to the general case.

\subsection{SII.A. The laser frequency is an integer multiple of the generalized Rabi frequency}
In this case, we start from the set of equations obtained in the main text for the Floquet harmonics $D_n$, $B_n$, and $Q_{n'}^{(n)}$ [see Eqs.~(13a), (13b), and (13c) of the main text]. To cast the eigenvalue problem in matrix form, we build a state vector with these harmonic coefficients, normalizing them so that all components share the same units and all matrix elements have units of frequency. To this end, we define
\begin{subequations}
    \begin{align}
        \tilde{D}_n &= \frac{D_n}{\varepsilon_0 c} , \\
        \tilde{Q}^{(n)}_{n'} &= \frac{\Omega_r}{c} Q^{(n)}_{n'} ,
    \end{align}
\end{subequations}
which have the same units as $B_n$. With these definitions, the set of equations in Eq.~(13) of the main text can be rewritten as
\begin{subequations} \label{Eq:eigenvalue_problem}
    \begin{align}
         \omega_F\tilde{D}_n &= n\Omega_r\tilde{D}_n + kc_0 B_n , \\
\omega_F B_n &= n\Omega_r B_n + kc_0\tilde{D}_n - ikc_0\sum_{n',p}\tilde{\varepsilon}_{n-p,n'}\tilde{Q}^{(n')}_p , \\
\omega_F\tilde{Q}^{(n)}_{n'} &= n'\Omega_r \tilde{Q}^{(n)}_{n'} + i\Omega_r\tilde{D}_{n'-n} + \Omega_r \sum_{p,p'}\tilde{\varepsilon}_{n'-p'-n,p}\tilde{Q}^{(p)}_{p'-n} ,
    \end{align}
\end{subequations}
where $\tilde{\varepsilon}_{n,n'}={\varepsilon}_{n,n'}/(\hbar \varepsilon_0 \Omega_r)$ is a dimensionless quantity.

The eigenvalue problem can then be written in matrix form as
\begin{equation}
    \overleftrightarrow{M}(k)\vec{u}=\omega_F \vec{u},
\end{equation}
where the vector is given by
\begin{equation}
   \vec{u} = \begin{pmatrix} \vec{\tilde{D}} \\ \vec{B} \\ \vdots \\ \vec{\tilde{Q}}^{(n)} \\ \vdots \\ \vec{\tilde{Q}}^{(p)} \\ \vdots \end{pmatrix},
\end{equation}
with $\vec{\tilde{D}}$ the column vector of normalized harmonic coefficients $\tilde{D}_n$, $\vec{B}$ the vector of harmonic coefficients $B_n$, and $\vec{\tilde{Q}}^{(n)}$, $\vec{\tilde{Q}}^{(p)}$ the vectors of harmonic coefficients $\tilde{Q}_{n'}^{(n)}$ and $\tilde{Q}_{n'}^{(p)}$, respectively; the ellipses indicate the concatenation over all auxiliary-variable indices. The matrix takes the form
\begin{equation}
    \overleftrightarrow{M} = \begin{pmatrix} \overleftrightarrow{\Lambda} & kc_0\,\overleftrightarrow{I} & \cdots & 0 & \cdots & 0 & \cdots \\ kc_0\,\overleftrightarrow{I} & \overleftrightarrow{\Lambda} & \cdots & -ikc_0\,\overleftrightarrow{Y}^{(n)} & \cdots & -ikc_0\,\overleftrightarrow{Y}^{(p)} & \cdots \\ \vdots & \vdots & \ddots & \vdots & & \vdots & \\ i\Omega\,\overleftrightarrow{X}^{(n)} & 0 & \cdots & \overleftrightarrow{\Lambda} + \Omega\,\overleftrightarrow{Z}^{(n,n)} & \cdots & \Omega\,\overleftrightarrow{Z}^{(n,p)} & \cdots \\ \vdots & \vdots & & \vdots & \ddots & \vdots & \\ i\Omega\,\overleftrightarrow{X}^{(p)} & 0 & \cdots & \Omega\,\overleftrightarrow{Z}^{(p,n)} & \cdots & \overleftrightarrow{\Lambda} + \Omega\,\overleftrightarrow{Z}^{(p,p)} & \cdots \\ \vdots & \vdots & & \vdots & & \vdots & \ddots \end{pmatrix} ,
\end{equation}
where we have introduced 
\begin{subequations}
    \begin{align}
        \overleftrightarrow{\Lambda}_{n,n''} = n\Omega\,\delta_{n,n''} , \\
        \overleftrightarrow{X}^{(n)}_{n',\,p} = \delta_{p,\,n'-n} ,\\
        \overleftrightarrow{Y}^{(n)}_{n'',\,p} = \tilde{\varepsilon}_{n''-p,\,n} , \\
        \overleftrightarrow{Z}^{(n,p)}_{n',\,p'} = \tilde{\varepsilon}_{n'-p'-n,\,p} .
    \end{align}
\end{subequations}

Figure 3 of the main text is obtained by solving numerically the eigenvalues of $\overleftrightarrow{M}$. We note that the state vector has $[\text{dim}(n)]^2 + 2\,\text{dim}(n)$ components, where $\text{dim}(n)$ is the number of Floquet harmonics retained in the truncated expansions of $B(t)$, $D(t)$ and $Q^{(n)}(t)$. Of these, only $2\,\text{dim}(n)$ correspond to physical fields ($D(t)$ and $B(t)$); the remaining $[\text{dim}(n)]^2$ components arise from the auxiliary variables $\tilde{Q}^{(n)}$ introduced to close the system of equations, and give rise to spurious eigenmodes with no physical meaning. To filter them out, we compute for each eigenmode the weight $f_{BD}$ of its eigenvector in the $(\tilde{D}, B)$ subspace:
\begin{equation}
  f_{BD} = \frac{\sum_{n}(|\tilde{D}_n|^2 + |B_n|^2)}{\sum_{n}(|\tilde{D}_n|^2 + |B_n|^2) + \sum_{n,n'}|\tilde{Q}^{(n)}_{n'}|^2}.
\end{equation}
Physical modes concentrate their weight in the $(\tilde{D}, B)$ subspace, while spurious modes have $f_{BD}$ close to zero. We retain only modes exceeding a threshold of $10^{-5}$ relative to the maximum value of $f_{BD}$ across all modes.

\subsection{SII.B. General case: incommensurate laser and Rabi frequencies}
In this section we derive the general eigenvalue problem without assuming that the laser frequency is an integer multiple of the generalized Rabi frequency. We start by substituting the time-domain polarizability [Eq.~(4) of the main text] into the polarization in Eq.~(7) of the main text, which yields
\begin{equation}
    P(t) = \frac{i }{\hbar}\sum_{m,m',l,l'} \varepsilon_{m,l,m',l'}e^{i(m\Omega_r +l\omega_L)t} Q^{(m',l')} (t) ,
\end{equation}
where we have defined $\varepsilon_{m,l,m',l'}=N_d \alpha_{m,l,m',l'}$, and the generalized auxiliary variables
\begin{equation} 
    Q^{(m,l)} (t) = \int_{-\infty}^\infty dt' \: \Theta(t-t')e^{i(m\Omega_r +l\omega_L)t} E(t'),
\end{equation}
which satisfy
\begin{equation} \label{Eq:differential_eq_Q_generalized}
    \frac{\partial Q^{(m,l)} (t)}{\partial t}=e^{i(m\Omega_r +l\omega_L)t}E(t).
\end{equation}

We next adopt Floquet ansatzes for the displacement field, the magnetic field, and the auxiliary variables. Since both modulation frequencies appear explicitly, these read
\begin{subequations}
    \begin{align}
        B(t) &= e^{-i\omega_F t}\sum_{p,q} B_{p,q}\,e^{i(p\Omega_r+q\omega_L)t} , \\
D(t) &= e^{-i\omega_F t}\sum_{p,q} D_{p,q}\,e^{i(p\Omega_r+q\omega_L)t},\\
Q^{(m,l)}(t) &= e^{-i\omega_F t}\sum_{p,q} Q^{(m,l)}_{p,q}\,e^{i(p\Omega_r+q\omega_L)t}.
    \end{align}
\end{subequations}
Substituting these expressions into the Faraday equation [Eq.~(6a) of the main text], the Ampère equation [Eq.~(6b) of the main text], and Eq.~(\ref{Eq:differential_eq_Q_generalized}), we obtain
\begin{subequations} \label{Eq:general_incomensurate_equations}
\begin{align}
    \omega_F D_{p,q} &= (p\Omega_r + q\omega_L)\,D_{p,q} + \frac{k}{\mu_0}B_{p,q} ,\\
    \omega_F B_{p,q} &= (p\Omega_r + q\omega_L)\,B_{p,q} + \frac{k}{\varepsilon_0}D_{p,q} - \frac{ik}{\varepsilon_0\hbar}\sum_{m,l,m',l'}\mathcal{E}_{m,l,m',l'}\,Q^{(m',l')}_{p-m,\,q-l} , \\
    \omega_F \sum_{p,q} Q^{(m,l)}_{p,q} &= \sum_{p,q}(p\Omega_r + q\omega_L)\,Q^{(m,l)}_{p,q} + \sum_{p,q}\frac{i}{\varepsilon_0}D_{p-m,\,q-l} + \frac{1}{\varepsilon_0\hbar}\sum_{p,q}\sum_{p',q'}\sum_{m',l'}\mathcal{E}_{p',q',m',l'}\,Q^{(m',l')}_{p-m-p',\,q-l-q'}.
\end{align}
\end{subequations}
We can then define the normalized quantities $\tilde{D}_{p,q} = D_{p,q}/(\varepsilon_0 c_0)$ and $\tilde{Q}^{(m,l)}_{p,q} = (\Omega_r/c_0)\,Q^{(m,l)}_{p,q}$, which have the same units as the magnetic field harmonic $B_{p,q}$, and solve the eigenvalue problem. Notably, Eqs.~(\ref{Eq:general_incomensurate_equations}) show that, in the general incommensurate case, the matrix elements couple harmonics through two independent indices, making the construction and diagonalization of the matrix significantly more complex than in the commensurate case (Section SII.A.). 

\section{SIII. Alternative Floquet eigenvalue problem with momentum as eigenvalue} \label{Appendix:Floquet_momentum}
In this section, we describe an alternative derivation of the Floquet eigenvalue problem, casting in this case the squared momentum $k^2$ as eigenvalue. To this end, we extend the formulation presented in Ref. \cite{Sloan_ACSPhotonics_2024} to the case with two modulation frequencies in the response function, which arise as we have explicitly included the modulation source.

In this case, we start from the wave equation for the electric field in the frequency domain, which can be written as
\begin{equation} \label{Eq:Wave-Equation-Freq}
\nabla^2 E(\omega)
+
\frac{\omega^2}{c^2}\frac{1}{2\pi}
\int d\omega' \biggr[1+ \frac{N_d}{\varepsilon_0} \alpha_{pr}(\omega,\omega')\biggr]E(\omega')
= 0,
\end{equation}
where we have assumed again a vacuum background for the emitters. Additionally, we have directly adopted a scalar description, since the transition dipole moment, pump field, and probe field are assumed to be linearly polarized along the same direction (e.g., in the $x$-direction).

We next propose the following Floquet ansatz for the electric field in the time domain,
\begin{equation}
    E(t)=e^{ikz}e^{-i\omega_F t}\sum_{M,L} E_{M,L}(\omega_F)e^{-iM\Omega_r t}e^{-iL\omega_L t},
\end{equation}
whose Fourier transform yields
\begin{equation}
    E(\omega)=e^{ikz}2\pi \sum_{M,L} {E}_{M,L}(\omega_F)\delta(\omega-\omega_F-M\Omega_r-L\omega_L).
\end{equation}
Substituting into Eq. (\ref{Eq:Wave-Equation-Freq}), the wave equation becomes
\begin{equation}
\begin{split}
    &\biggr[k^2 - \frac{(\omega_F+M\Omega_r+L\omega_L)^2}{c^2}\biggr]{E}_{M,L}(\omega_F )\\
    &-\frac{(\omega_F+M\Omega_r+L\omega_L)^2}{c^2}\frac{N_d}{\varepsilon_0}\sum_{M',L'} \alpha_{M',L'} (\omega_F+M\Omega_r+L\omega_L){E}_{M-M',L-L'}(\omega_F )=0 .
\end{split}
\end{equation}

Performing the index shifts $M'\rightarrow M-M'$ and $L'\rightarrow L-L'$, we obtain
\begin{equation}
\begin{split}
    &\biggr[k^2 - \frac{(\omega_F+M\Omega_r+L\omega_L)^2}{c^2}\biggr]{E}_{M,L}(\omega_F )\\
    &-\frac{(\omega_F+M\Omega_r+L\omega_L)^2}{c^2}\frac{N_d}{\varepsilon_0}\sum_{M',L'} \alpha_{L-L',M-M'} (\omega_F+M\Omega_r+L\omega_L){E}_{M',L'}(\omega_F )=0 .
\end{split}
\end{equation}
This equation can be rearranged as
\begin{equation}
\begin{split}
    &\sum_{M',L'}\frac{(\omega_F+M\Omega_r+L\omega_L)^2}{c^2}\biggr(\delta_{l,l'}\delta_{m,m'}+\frac{N_d}{\varepsilon_0}\alpha_{M-M',L-L'} (\omega_F+M\Omega_r+L\omega_L)\biggr){E}_{M',L'}(\omega_F )=k^2 {E}_{M,L}(\omega_F ).
\end{split}
\end{equation}

By reindexing $(M,L)\rightarrow n$ through a one-to-one mapping, the above equation can be cast as a matrix eigenvalue problem. Specifically, we truncate the $L$-manifold to dimension $\text{dim}(L)=2N_L+1$, so that $L\in\{-N_L,-N_L+1,...,0,...,N_L,N_L+1\}$, and similarly the $M$-manifold to dimension $\text{dim}(M)=2N_M+1$, so that $M\in\{-N_M,-N_M+1,...,0,...,N_M,N_M+1\}$. The combined $n$-manifold then has dimension $\text{dim}(n)=(2N_L+1)(2N_M+1)=4N_L N_M +2N_L +2N_M + 1=2N_n + 1$, with $N_n=2N_L N_M + N_L + N_M$, and the index $n$ runs over $0$, $1$, $2$, ..., $N_n-1$. The mapping from $(M,L)$ to $n$ is explicitly given by 
\begin{equation}
n=(l+N_L)(2N_M+1)+M+N_M.    
\end{equation}
The eigenvalue problem then takes the matrix form
\begin{equation}
    [\overleftrightarrow{T}(\omega_F)]\vec{E}(\omega_F+\omega_L)=k^2 \vec{E}(\omega_F),
\end{equation}
where the matrix elements are 
\begin{equation}
    \overleftrightarrow{T}_{nn'}(\omega_F)=\frac{(\omega_F+M\Omega_r+L\omega_L)^2}{c^2}\biggr(\delta_{L,L'}\delta_{M,M'}+\frac{N_d}{\varepsilon_0}\alpha_{M-M',L-L'} (\omega_F+M\Omega_r+L\omega_L)\biggr).
\end{equation}

This formulation yields a smaller matrix problem than that obtained in Section SII, where the eigenvalue corresponds to $\omega_F$, since it only requires solving for the harmonics of the electric field rather than those of the displacement field, the magnetic field, and all auxiliary variables, resulting in significantly shorter computation times. However, the electromagnetic modes obtained within the formulation with $k^2$ as eigenvalue include both evanescent and propagating solutions, making it numerically difficult to isolate the latter. This complicates the physical analysis of the band structure and, in particular, that of the dispersive momentum bandgaps. Nevertheless, we have verified that the results obtained with this formulation are consistent with those presented in the main text. Moreover, the calculation of the phase rigidity and the identification of the exceptional points (EPs) are naturally obtained by constructing the eigenvalue problem in $\omega_F$ rather than in $k^2$. Indeed, EPs correspond to values of $k$ at which two eigenvalues $\omega_F$ coalesce together with their associated eigenvectors, so that the phase rigidity vanishes. When the eigenvalue problem is instead formulated in $k^2$ for fixed $\omega_F$, such coalescence does not appear directly as a degeneracy of the $k^2$ eigenvalues and requires additional analysis to identify.

\section{SIV. Band structure for different emitter densities}
\begin{figure*}[!t]  
	\begin{center}
		\includegraphics[width=0.9\textwidth]{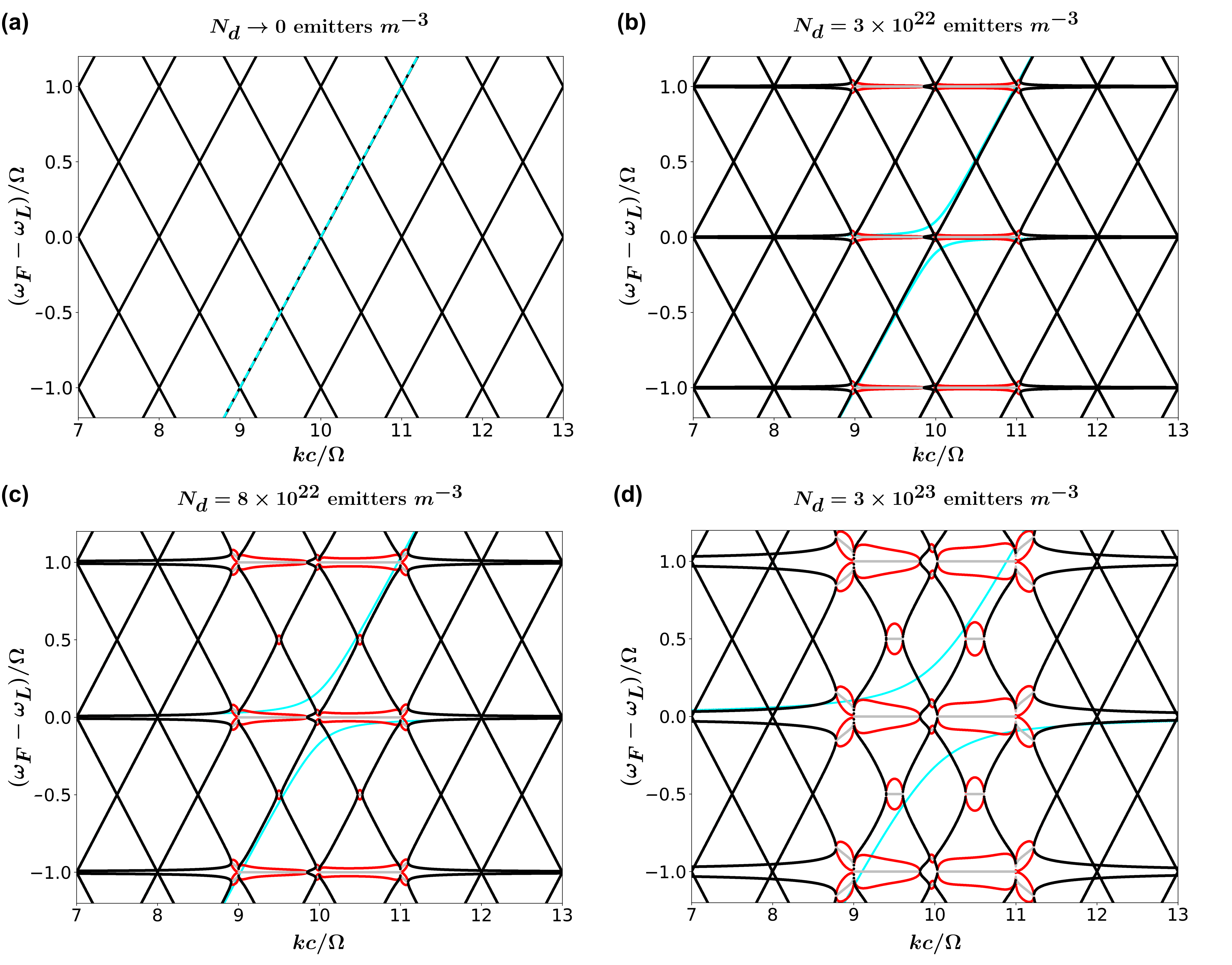}%
		\caption{Dispersion relation for an ensemble of quantum emitters in the Mollow regime and different values of emitter density. We fix $\Omega=\Omega_r=\omega_0 /10$, and $\hbar\omega_0 = 1$ eV, as in the main text. The emitter density is fixed at (a) $\rightarrow0$ emitters per $\text{m}^{3}$, (b) $N_d=3\times10^{22}$ emitters per $\text{m}^{3}$, (c) $N_d=8\times10^{22}$ emitters per $\text{m}^{3}$, and (d) $N_d=3\times10^{23}$ emitters per $\text{m}^{3}$, the same value as in Fig. 3 of the main text.   Modes with purely real Floquet frequency are represented by black dots, with the $y$-axis value corresponding to $(\text{Re}\{\omega_F\}-\omega_L)/\Omega$. Modes with a complex Floquet frequency are represented by grey and red dots: the grey dots show $(\text{Re}\{\omega_F\}-\omega_L)/\Omega$, while the red dots show the value of $\text{Im}\{\omega_F\}/\Omega$ shifted by $(\text{Re}\{\omega_F\}-\omega_L)/\Omega$. Cyan dashed line in (a) represent the light line dispersion relation. Cyan dots in [(b)-(d)] correspond to the polaritonic dispersion relation in the absence of a pump.} \label{Fig:AppBands}
	\end{center}
\end{figure*}

In this section, we analyze the emergence of the band structure in Figs. 2 and 3 of the main text as a function of the modulation strength. To this end, we investigate the band structure for different values of emitter density, from $N_d\rightarrow0$ emitters per m$^3$ to the value $N_d=3\times10^{23}$ emitters per m$^{3}$ in Fig. 3 of the main text. The emitter and laser parameters are the same as those considered in the main text, i.e., $\Omega=\Omega_r=\omega_L /10$, and $\hbar\omega_0 = 1$ eV. Additionally, to solve the eigenvalue problem numerically, we use the formulation discussed in Section SII.A of this Supplemental Material, which is the same formulation used in the calculation of the band structure in Figs. 2 and 3 of the main text. The color convention for black, grey, and red dots in Fig. \ref{Fig:AppBands} is the same as that adopted in Fig. 2 of the main text.

We start by setting $N_d\rightarrow0$ emitters per m$^3$ in Fig. \ref{Fig:AppBands}(a), which is equivalent to setting a vanishing modulation strength. The permittivity thus reduces to the vacuum background permittivity and the dispersion relation to the vacuum light line $k=\omega_F/c$ (dashed cyan line). Within the periodic lattice approximation, this relation is replicated at the harmonic modulation frequencies \cite{Galiffi_Review_2022, Carsten_tutorial_2024}, leading to the dispersion relation shown by the black dots in Fig. \ref{Fig:AppBands}(a), obtained by solving the Floquet eigenvalue problem in Eq. (13) of the main text.

We next increase the emitter density to a value where small momentum bandgaps emerge. Specifically, we fix $N_d=3\times10^{22}$ emitters per m$^{3}$ in Fig. \ref{Fig:AppBands}(b), a value ten times smaller than that in Figs. 2 and 3 of the main text. In Fig. \ref{Fig:AppBands}(b) we additionally plot the polaritonic dispersion relation obtained in the absence of pumping (cyan dots), which is given by $k = \frac{\omega_F}{c}\sqrt{1+\frac{N_d}{\varepsilon_0}\alpha_0(\omega)}$, with $\alpha_0(\omega)=\frac{d^2}{\hbar}\!\left[\frac{1}{\omega+\omega_0}-\frac{1}{\omega-\omega_0}\right]$, and which reduces to the light line in Fig. \ref{Fig:AppBands}(a) when $N_d\rightarrow0$. We observe that at $N_d=3\times10^{22}$ emitters per m$^{3}$ a small polaritonic Rabi splitting already emerges in Fig. \ref{Fig:AppBands}(b), leading to the flat bands shown by the black dots in the outer $k$ region ($kc/\Omega<9$ and $kc/\Omega>10$). Additionally, we observe flat momentum bandgaps at the frequencies of the Mollow triplet peaks, i.e., at $\omega_F = \omega_L \pm \Omega$ and $\omega_F = \omega_L$. However, this value of emitter density (equivalently, modulation strength) is not sufficient to clearly observe the flat momentum bandgaps at $\omega_F = \omega_L \pm \Omega/2$, nor the dispersive momentum bandgaps discussed in the main text. 

We next fix $N_d=8\times10^{22}$ emitters per m$^{3}$ in Fig. \ref{Fig:AppBands}(c). The polaritonic Rabi splitting obtained in the absence of the pump (cyan dots) increases, and the slope of the bands obtained in the pumped case in the outer $k$-region slightly increases with respect to Fig. \ref{Fig:AppBands}(b). Additionally, the momentum bandgaps at $\omega_F = \omega_L \pm \Omega/2$ and the dispersive momentum bandgaps start to emerge. However, the flat momentum bandgaps at the frequencies of the Mollow triplet peaks again exhibit a larger imaginary part of the Floquet frequency (red dots).

Finally, for completeness, we plot in Fig. \ref{Fig:AppBands}(d) the dispersion relation at the same emitter density $N_d=3\times10^{23}$ emitters per m$^{3}$ as in Figs. 2 and 3 of the main text, where the momentum bandgaps are clearly visible. Thus, Fig. \ref{Fig:AppBands}(d) corresponds to Fig. 2 of the main text.

\bibliography{bibliography.bib}